\documentclass{PoS}
\usepackage{epsfig}
\PoS{PoS(LAT2005)143}
\newcommand{\beq}{\begin{equation}}
\newcommand{\eeq}{\end{equation}}
\newcommand{\sign}{\mathop{\rm sign}}

\newcommand{\MeV}{{\mbox{MeV}}}

\title{Localization of Low Lying Eigenmodes for Chirally Symmetric Dirac
Operator
} \ShortTitle{Localization of Low Lying Eigenmodes for Chirally
Symmetric Dirac Operator
 }
\author{\speaker{M.I. Polikarpov}, F.V.~Gubarev, S.M.~Morozov and
\\
ITEP, B. Cheremushkinskaya 25, Moscow, 117259 Russia\\
        E-mail: \email{polykarp@itep.ru}, \email{gubarev@itep.ru},
        \email{smoroz@itep.ru}}
\author{V.I.~Zakharov\\
 MPI, F\"ohringer Ring 6, 80805, M\"unchen, Germany\\
        E-mail: \email{xxz@mppmu.mpg.de}}
\abstract{We consider properties of zero and near-zero modes for
overlap fermion operator in SU(2) lattice gluodynamics. The density of
the states is of the order of $\Lambda_{QCD}$ while the localization
volume of the modes tends to zero in physical units with the lattice
spacing tending to zero. The situation changes drastically when we
study "vortex removed" configurations.}
\FullConference{XXIIIrd International Symposium on Lattice Field Theory\\
         25-30 July 2005\\
         Trinity College, Dublin, Ireland}
\begin{document}
\section{Introduction}
The material of this talk has a substantial overlap with that presented
in the talks of N.~Cundy, C.~Gattringer, T.~De~Grand, J.~Greensite,
J.~Hetrick, I.~Horvath, Y.~Koma, S.~Solbrig, B.~Svetitsky, S.~Syritsyn.

We study SU(2) lattice gluodynamics and in our calculations we use the
massless overlap Dirac operator~\cite{neuberger}:
\beq
\label{overlap} D = \frac{\rho}{a} \left( 1 +
\frac{A}{\sqrt{AA^\dagger}} \right),\,\quad A = D_W - \frac{\rho}{a},
\end{equation}
where $A$ is the Wilson Dirac operator with negative mass term.
Anti-periodic (periodic) boundary  conditions in time (space)
directions were employed. It turns out that for SU(2) gluodynamics the
optimal value of $\rho$ parameter is $1.4$.  Furthermore, we have used
the minmax polynomial approximation~\cite{giusti} to compute the sign
function $\sign(A) = A/\sqrt{AA^\dagger} \equiv \gamma_5 \sign(H)$,
where $H=\gamma_5 A$ is hermitian Wilson Dirac operator. In order to
improve the accuracy and performance about one hundred lowest
eigenmodes of $H$ were projected out. Note that the eigenvalues of
(\ref{overlap}) lie on the circle of radius $\rho$ centered at
$(\rho,0)$ in the complex plane. In order to relate them with
continuous eigenvalues of the Dirac operator the circle was
stereographically projected onto the imaginary axis~\cite{capitani}.
The information about statistics and number of gauge field
configurations used can be found in~\cite{morozovjetp}.

Below we discuss localization properties of eigenmodes of the
overlap Dirac operator. A natural measure of the
localization is provided by the inverse participation ratio (IPR)
$I_{\lambda}$ which is defined as follows (for review see for example
Ref.~\cite{CM}). Let
$$
\rho_\lambda(x) = \psi_\lambda^\dagger(x)\psi_\lambda(x)\,,\qquad
\sum_x \rho_\lambda(x) = 1\,,
$$
where $\psi_{\lambda}(x)$ is an eigenmode of the overlap Dirac
operator in a given gauge field background with virtuality $\lambda$,
$D\,\psi_{\lambda} = \lambda\,\psi_{\lambda}$. Then for any finite
volume $V$ the IPR $I_{\lambda}$ is defined by
\beq
\label{eq:ipr_def} I_\lambda = V \, \sum_x \rho_\lambda^2(x)\,,
\end{equation}
and characterizes the inverse fraction of sites contributing to the
support of $\rho_{\lambda}(x)$. Note that for delocalized modes
$\rho_{\lambda}(x) = 1/V$ and hence $I_{\lambda} = 1$, while an
extremely localized mode, $\rho_{\lambda}(x) = \delta_{x,x_0}$, is
characterized by $I_\lambda = V$. Moreover, for eigenmodes localized on a
fraction $f$ of sites (so that $\mathrm{sup}~\rho_\lambda = V_f = f ~
V$) we have $I_{\lambda} = V/V_f = 1/f$.

\section{Dependence of the IPR on the eigenvalue}
\begin{figure}
\centerline{\psfig{file=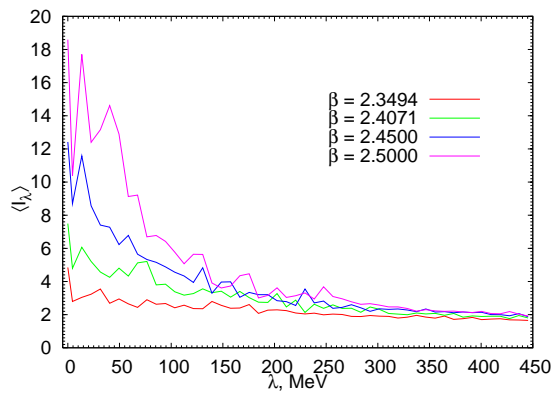,width=0.5\textwidth,silent=,clip=}}
\caption{IPRs for low lying eigenmodes at various lattice spacings and
fixed physical volume. The ``mobility edge'' $\lambda_{cr} \approx 150
- 200\,\MeV$ is clearly seen.} \label{ipr_histo_a}
\end{figure}
\begin{figure}
\centerline{\psfig{file=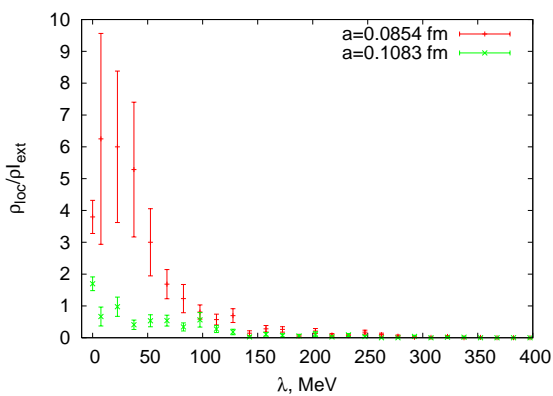,angle=-90,width=0.5\textwidth,silent=,clip=}}
\caption{The ratio of the densities of the eigenmodes with IPR$>5$ to
that with IPR$<5$.} \label{loc_ext}
\end{figure}
In Fig.~\ref{ipr_histo_a} we show IPR for various lattice spacings at
fixed physical volume. Two different effects are clearly observed.
First, the IPR grows as function of $\lambda_n$ with $\lambda_n\to 0$,
for fixed total volume and fixed lattice spacing. Second, for
sufficiently small, fixed  values of $\lambda_n$ the IPR grows with the
lattice spacing tending to zero. An alternative way of representing the
growth of IPR with $a\to 0$ is provided by the data on
Fig.~\ref{loc_ext}. Here we plot the ratio, $R$ of the densities of the
eigenmodes with IPR$>5$ and with IPR$<5$ for two values of the lattice
spacing $a$. The data demonstrate that the observed growth of the IPR
with $a\to 0$ is due to `typical' eigenmodes, not due to a few modes
with a huge value of the IPR.

Another qualitative feature of the data is existence of a kind of a
`merging point', $\lambda_{merge} \approx 150 - 200\,\MeV$. Namely, for
$\lambda_n > \lambda_{merge}$ the values of IPR practically do not
depend  on the eigenvalue $\lambda_n$ and the lattice spacing $a$. Note
that existence of a point where all the curves giving the dependence of
the IPR on $\lambda_n$ is also seen in the data on localization
properties of eigenfunctions of a test color scalar particle
\cite{olejnik}. Our data indicate scaling in physical units, at least
an approximate one, of the position of the merging point.

Thus, our data on the localization properties of the low-lying
eigenfunctions of the Dirac operator indicate existence of a variety of
phenomena. If we look for analogies into solid-state physics (reviewed
in, e.g., \cite{CM}) then, probably, one can find an analog to the
growth of the IPR as function of $\lambda_n$ with diminishing
$\lambda_n$. Indeed, it is known that if one approaches the mobility
edge, $\lambda_{cr}$~\footnote{Let us remind the reader that the
mobility edge is defined in terms of dependence of the IPR on the total
volume of the system, $V_{tot}$. Namely, $(IPR)\sim V_{tot}$ for
$V_{tot}\to\infty$ below the mobility edge, $\lambda_n <\lambda_{cr}$
and $(IPR)\sim const$ for $V_{tot}\to \infty$ above the mobility edge,
$\lambda_n <\lambda_{cr}$. Our data on the dependence of the IPR on the
total volume can be found in the original paper \cite{morozovjetp}.}
from above,
 then  the IPR blows up as $(\lambda_n-\lambda_{cr})^{-\alpha}$ where $\alpha >0$.
Note that this simple dependence holds for a fixed, large total volume
$V_{tot}$ and the singularity is smoothened due to finiteness of the
total volume. Now, for chiral fermions one expects that the mobility
edge is $\lambda_{cr}=0$. Then, the growth of the IPR with
$\lambda_n\to 0$ indicated by our data (see Fig.~\ref{ipr_histo_a}) is
reminiscent of existence of the critical exponent $\alpha$ in the
non-relativistic case.

Since the data exhibit strong dependencies of the IPR on various
variables, detailed analysis  of these dependencies goes beyond the
scope of the present study (some details, though, can be found  in the
original paper \cite{morozovjetp}). Here, we confine ourselves mostly
to discussion of qualitative effects. From this point of view
dependence on the ultraviolet cut off, that is the lattice spacing $a$,
seems to be most exciting. We will discuss this point further in Sect.
4.

\section{Removing of central vortices}

\begin{figure}
\centerline{\psfig{file=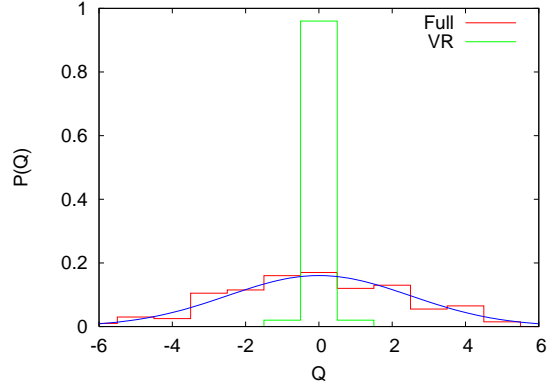,angle=-90,width=0.5\textwidth,silent=,clip=}}
\caption{The histogram of the distribution of the topological charge
for full and vortex removed (VR) configurations.} \label{pQ_full_vr}
\end{figure}

The vortex-removing procedure~\cite{vortrem} drastically changes the
properties of the  vacuum: confinement and chiral condensate
disappear. Below we describe how the vortex removing acts on topological
susceptibility of the vacuum, distribution of the Dirac operator
eigenvalues and localization properties of the eigenmodes. On
Fig.\ref{pQ_full_vr} we show the distribution of the topological charge
for full and vortex-removed configurations. It occurs that topological
susceptibility, $<Q^2/V>$, vanishes after removing of the center
vortices. The topological charge of the configuration was defined by
the index of overlap Dirac operator $Q = n_+ - n_-$ and $n_\pm$ is the
number of exact zero modes with positive (negative) chirality.

\begin{figure}
\centerline{\psfig{file=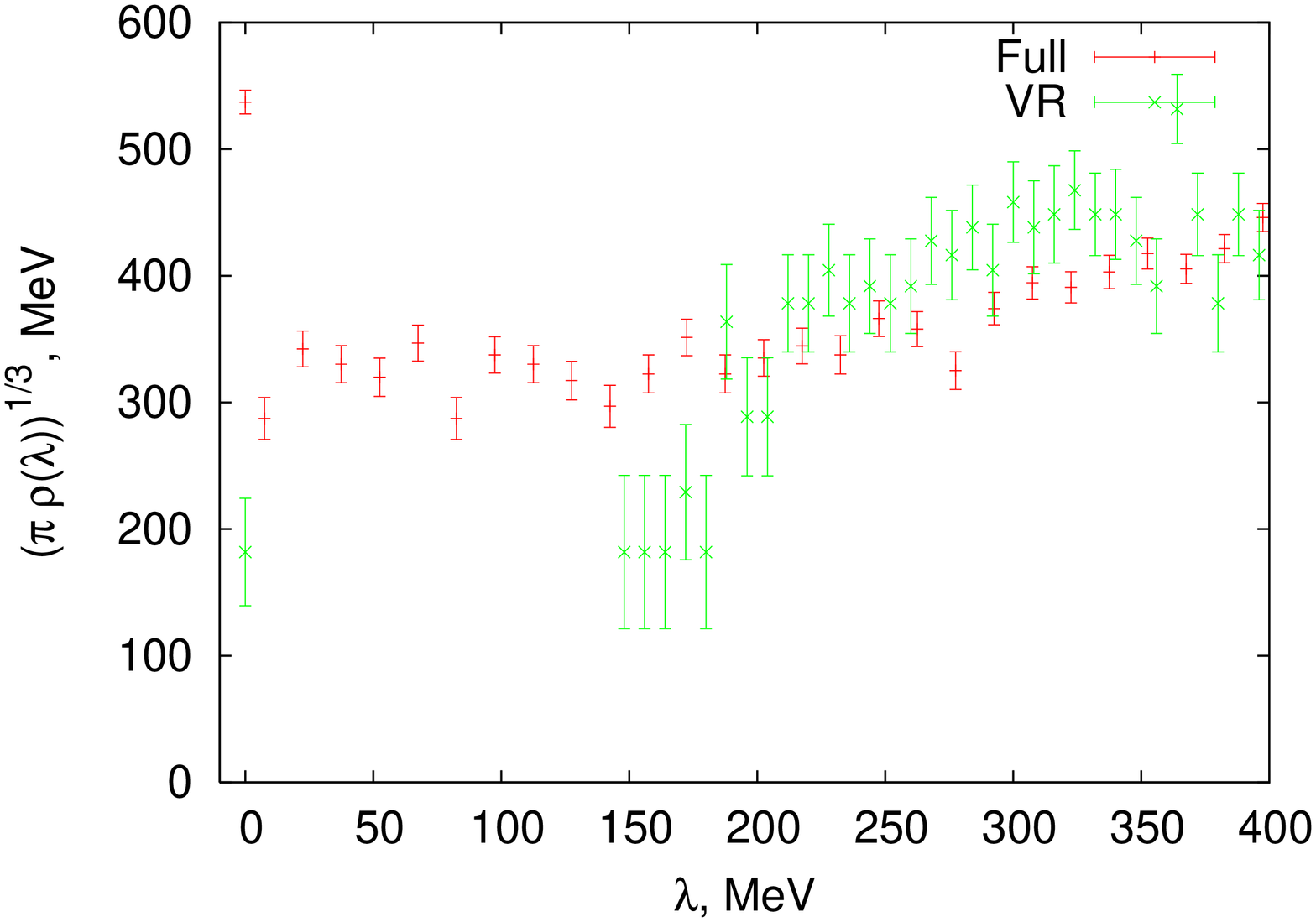,angle=-90,width=0.5\textwidth,silent=,clip=}
\psfig{file=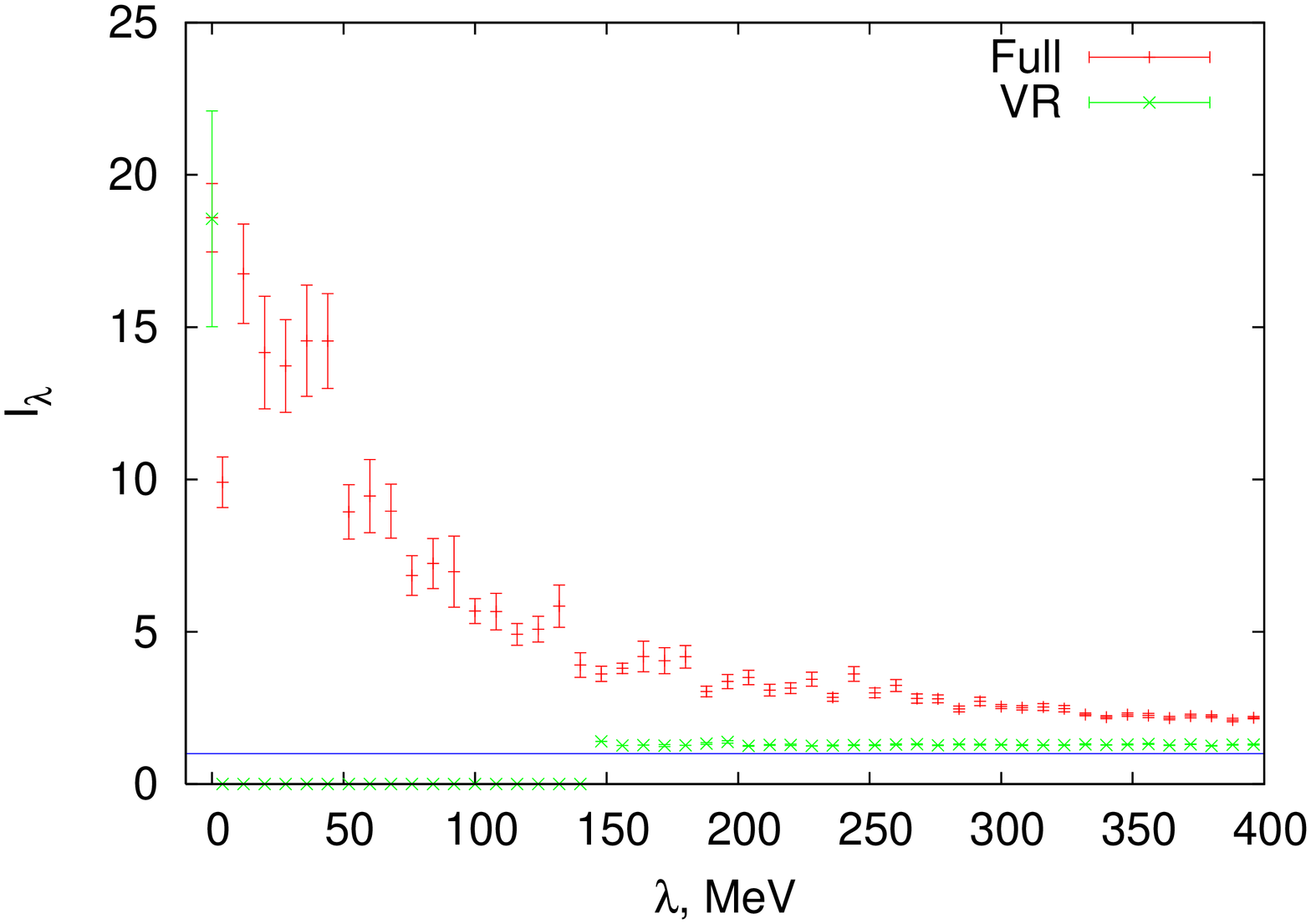,angle=-90,width=0.5\textwidth,silent=,clip=}}
\caption{$<\pi\rho(\lambda)>^\frac13$ (left) and IPR (right) for full
and vortex removed (VR) configurations. The average is taken over 200
gauge field configurations at $\beta=2.5$ on lattice $16^3\cdot 18$.}
\label{rho_full_vr}
\end{figure}

On Fig.\ref{rho_full_vr} (left) we show the quantity
$<\pi\rho(\lambda)>^\frac13$ which is related to chiral condensate
through the Banks-Casher relation \cite{banks-casher}, $\langle
\bar{\psi}\psi\rangle=-\pi \rho(\lambda_n\to 0)$, where
$\rho(\lambda_n\to 0)$ is the density of the (delocalized) zero modes
in the limit of infinite volume $V$. For finite lattice volumes these
modes are near-zero and this is the reason of the limit $\lambda_n\to
0$.
It is seen that for the vortex removed configurations there exist the
gap in the distribution of the near-zero eigenvalues, thus chiral
condensate vanishes for vortex removed gauge fields.

On Fig.\ref{rho_full_vr} (right) IPR for full and vortex removed
configurations is shown. After vortex removing all modes (except of a
few surviving zero modes) become delocalized, $(IPR)\approx 1$.

Thus, we can conclude that the non-trivial localization properties of the
eigenfunctions of the Dirac operator are directly related to the central vortices.
Note that similar observations were made by using other types of lattice fermions, see
in particular \cite{gattnar} and references therein.

\section{Discussions}

The most striking feature exhibited by the data is a strong dependence of the IPR
on the lattice spacing $a$. The data were fitted \cite{morozovjetp} by
\beq
I_{\lambda}~=~b_0~+~b_1\cdot a^{d-4}~~,
\end{equation}
where $b_{0,1},d$ are constants to be determined from the fit.
The IPR data for the low lying modes strongly suggest that
\beq
d~\approx~0~~.
\end{equation}
In other words, the volume occupied by the eigenfunctions tends to zero
in physical units as $a^4$.

Note that the $a$ dependence of the IPR was suggested first by the data of Ref. \cite{aubin}.
The lattice-spacing dependence exhibited by our data is considerably stronger
than that reported in Ref. \cite{aubin}. In particular, in Ref. \cite{aubin} the
values of IPR were changing, say, between $(IPR)\approx 2$ and $(IPR)\approx 3$
as a result of varying the lattice spacing. In our case, the highest values of IPR are
approaching the value of 20. At this moment, it is difficult to clarify the reason
for this difference. First, we use different types of fermions. Also, the color
group is different, as well as the lattice action used.

The strong $a$ dependence of the IPR is in sharp contrast with scaling,
in physical units, of the density of states and of the chiral
condensate (defined through the Casher-Banks relation). Thus, the
emerging picture is that the size of the eigenfunctions shrinks
strongly with $a\to 0$ while the eigenvalues remain stable. In other
words, the data suggest that the phenomenon of fine tuning is quite a
common feature of the Yang-Mills dynamics \footnote{The standard
definition of the fine tuning, which goes back to the Higgs physics, is
as follows. If a relativistic system has a typical size $r_0$ then the
energy levels of order $\lambda \sim 1/r_0$ are `natural'. If, on the
other hand, $\lambda \ll 1/r_0$ the level is `fine tuned'.}.

Note that the phenomenon of the fine tuning was observed first by studying the
properties of the lattice monopoles and vortices, for a review see, e.g., \cite{viz}.
However the definition of the monopoles and vortices involves projected fields
which makes theoretical interpretation more difficult. The fine tuning seen
in the data on the fermion localization is defined in explicitly gauge invariant terms.

\begin{acknowledgments}
The invaluable assistance of G.~Schierholz and T.~Streuer in the
overlap operator implementation is kindly acknowledged. This work was
partially supported  by grants RFBR-05-02-16306a, RFBR-05-02-17642,
RFBR-0402-16079, RFFI-05-02-17642, RFBR-03-02-16941 and  EU Integrated
Infrastructure Initiative Hadron Physics (I3HP) under contract
RII3-CT-2004-506078. F.V.G. was partially supported by INTAS YS grant
04-83-3943.
\end{acknowledgments}

\end{document}